\documentclass[final,3p,times]{elsarticle}

\usepackage[utf8]{inputenc}
\usepackage{geometry}
\usepackage{url}
\usepackage{amssymb}
\usepackage{lineno}

\usepackage[autolinebreaks,useliterate]{mcode}

\begin{document}

\begin{frontmatter}

\title{Exact distribution of selected multivariate test criteria by numerical inversion of their characteristic functions}

\author{Viktor Witkovský\corref{cor1}}
\ead{witkovsky@savba.sk}

\cortext[cor1]{Corresponding author}

\address{Institute of Measurement Science, Slovak Academy of Sciences, Dúbravská cesta 9, Bratislava, Slovakia}

\journal{ArXiv (\url{arxiv.org/list/math.ST})}

\begin{abstract}
Application of the exact statistical inference frequently leads to a non-standard probability distributions of the considered estimators or test statistics. The exact distributions of many estimators and test statistics can be specified by their characteristic functions. Typically, distribution of many estimators and test statistics can be structurally expressed as a linear combination or product of independent random variables with known distributions and characteristic functions, as is the case for many standard multivariate test criteria. The characteristic function represents complete characterization of the distribution of the random variable. However, analytical inversion of the characteristic function, if possible, frequently leads to a complicated and computationally rather strange expressions for the corresponding distribution function (CDF/PDF) and the required quantiles. As an efficient alternative, here we advocate to use the well-known method based on numerical inversion of the characteristic functions --- a method which is, however, ignored in popular statistical software packages. The applicability of the approach is illustrated by computing the exact distribution of the Bartlett's test statistic for testing homogeneity of variances in several normal populations and the Wilks's $\Lambda$-distribution used in multivariate hypothesis testing.
\end{abstract}


\end{frontmatter}


\section{Introduction}
In 1937, Bartlett proposed a testing procedure to test the hypothesis of equal variances of $k$ normal populations. He suggested an ingenious correction of the modified likelihood ratio based test statistic which under the null hypothesis approximately follows the chi-squared distribution with $\nu = k-1$ degrees of freedom, even for small sample sizes, see \cite{Bartlett1937}. In fact, the Bartlett-type corrections (multiplying factors) are known to be effective and precise for various approximate tests based on the asymptotic approximations for likelihood ratios for wide range of parameters (as e.g. the number of normal populations $k$ and the sample sizes $n_i$ for $i = 1,\dots,k$ in testing the homogeneity of variances), \cite{Bartlett1954,Cribari1996}. However, detailed comparison with the exact distribution is still desirable. 

In general, application of the exact statistical inference leads to a non-standard probability distributions of the considered estimators or test statistics which can be specified by their characteristic functions (CFs). Frequently, distribution of many estimators and test statistics can be structurally expressed as a linear combination or product of independent random variables with known characteristic functions, as is the case for many standard multivariate test criteria, see, e.g., \cite{Mathai1973,Arnold2013}. In such cases, analytical expressions for the exact distributions are typically difficult to derive. Hence, such distributions are usually approximated by using results of the asymptotic theory, see \cite{Mardia1979,Anderson2003}, or other available small sample approximation/correction methods, and frequently by using computer intensive simulation methods. 

In this paper we advocate using numerical inversion of the known characteristic function as an efficient tool to evaluate the required distribution --- the probability density function (PDF), as well as the cumulative distribution function (CDF), and the quantiles of such estimators or test statistics. The method based on numerical inversion is convenient when the characteristic function is known apriori and is such that it can be easily evaluated by the available algorithms, or if the statistic under consideration is a linear combination of independent random variables with known and simple characteristic functions. In order to apply the method to products of independent random variables one has to consider, first, the logarithm of the statistic and, subsequently, to transform the computed values to the original scale (if required). Unfortunately, such numerical algorithms are not available in standard statistical software packages (e.g.~SAS, R, MATLAB).

For illustration (and as a gentle introduction to the problem) let us consider first the distribution of a quadratic form $Q = X'AX$ with $X \sim N_n(0,\Sigma)$, where $\Sigma$ denotes the known covariance matrix and $A$ denotes the known p.s.d.~matrix of the quadratic form, then
\begin{equation}\label{eq01}
	Q \sim \sum_{j=1}^n \lambda_j Q_j,
\end{equation}
where $\lambda_j$ are eigenvalues of $A\Sigma$ and $Q_j \sim \chi^2_1$ are independent chi-square distributed random variables (RVs) with 1 degree of freedom. Hence, the characteristic function of $Q$, say $\mathop{\mathrm{cf}}\nolimits_{Q}(t)$, is given by
\begin{equation}\label{eq02}
\mathop{\mathrm{cf}}\nolimits_{Q}(t) = \prod_{j=1}^n \mathop{\mathrm{cf}}\nolimits_{Q_j}(\lambda_j t) = \prod_{j=1}^n  \left(1 - 2\mathrm{i} \lambda_j t \right)^{-\frac{1}{2}},
\end{equation}
where $\mathrm{i} = \sqrt{-1}$ denotes the imaginary unit and $\mathop{\mathrm{cf}}\nolimits_{Q_j}(t)  = \left(1 - 2\mathrm{i} t \right)^{-\frac{1}{2}}$, for all $j=1,\dots,n$, is CF of the chi-square distribution with 1 degree of freedom. The cumulative distribution function of $Q$, say $\mathop{\mathrm{cdf}}\nolimits_{Q}(x) = \Pr(Q\leq x)$, is a non-standard distribution (in general, the closed form expression is unknown), however, it can be evaluated numerically from its CF, as it was suggested in \cite{Imhof1961,Davies1980}.

This can be naturally generalized for more complicated applications, e.g., based on Gaussian stochastic processes. For example, let us consider the (asymptotic) distribution of the Cramér-von Mises and the Anderson-Darling statistics. These statistics belong to the class of quadratic goodness-of-fit test statistics based on the empirical distribution function. 
By using the theory of stochastic processes, the asymptotic distributions are derived from the Karhunen-Loève representation of functionals of the Brownian bridge. 

In particular, let $\hat{F}_n(x)$ denotes the empirical CDF based on $n$ i.i.d.~random variables $X_1,\dots,X_n$ from continuous distribution $F$, i.e.~$X_j \sim F$. Then, for $n\rightarrow \infty$, the distribution of the Cram\'er-von Mises statistic $W_n$ converges to the distribution of infinite sum of (weighted) independent chi-square distributed random variables with 1 degree of freedom, i.e.
\begin{equation}\label{eq07}
W_n = n\int_{-\infty}^\infty \left(\hat{F}_n(x) - F(x)\right)^2\,dF(x) \stackrel{L}{\longrightarrow}
W_\infty = \int_0^1 B^2(t)\,dt \sim \sum_{j=1}^\infty \frac{1}{(j\pi)^2}Q_j,
\end{equation}
where $B(t)$ represents the Brownian bridge process and $Q_j \sim \chi^2_1$ are i.i.d.~RVs. The exact distribution of $W_\infty$ is difficult to derive and evaluate, however its characteristic function is rather simple,
\begin{equation}\label{eq04}
\mathop{\mathrm{cf}}\nolimits_{W_\infty}(t) =  \prod_{j=1}^\infty \mathop{\mathrm{cf}}\nolimits_{Q_j}\left(\textstyle \frac{t}{(j \pi)^2}\right)  
= \prod_{j=1}^\infty \left(\textstyle 1-\frac{2 \mathrm{i} t}{(j\pi)^2}\right)^{-\frac12}  
= \sqrt{{\frac{\sqrt{2 \mathrm{i} t}}{\sin\left(\sqrt{2 \mathrm{i} t}\right)}}}.
\end{equation}

Similarly, for $n\rightarrow \infty$, the distribution of the Anderson-Darling statistic $A_n$ converges to the distribution of infinite sum of (weighted) independent chi-square distributed random variables with 1 degree of freedom, i.e.
\begin{equation}\label{eq05}
	A_n = n\int_{-\infty}^\infty \frac{\left(\hat{F}_n(x) - F(x)\right)^2}{F(x)(1-F(x))}\,dF(x)
	\stackrel{L}{\longrightarrow}  A_\infty =  \int_0^1 \frac{B^2(t)}{t(1-t)}\,dt \sim \sum_{j=1}^\infty \frac{1}{j(j+1)}Q_j,
\end{equation}
with its (rather simple) characteristic function given by
\begin{equation}\label{eq06}
\mathop{\mathrm{cf}}\nolimits_{A_\infty}(t)  =  \prod_{j=1}^\infty \mathop{\mathrm{cf}}\nolimits_{Q_j}\left(\textstyle\frac{t}{j(j+1)}\right)  
= \prod_{j=1}^\infty \left(\textstyle 1-\frac{2 \mathrm{i} t}{j(j+1)}\right)^{-\frac12} 
=  \sqrt{ \frac{-2\pi \mathrm{i} t}{\cos\left(\frac{\pi}{2}\sqrt{1+8\mathrm{i} t}\right)}}.
\end{equation}
For more details see \cite{Anderson1952,Marsaglia2004}. The distribution functions (PDF/CDF) of  $W_\infty$ and  $A_\infty$ can be evaluated numerically from their respective CFs by using the proper numerical inversion algorithm. 

The rest of the paper is organized as follows: In Section~\ref{BartlettsCF} we present the exact CF of the Bartlett's $\chi^2$ test statistic for testing homogeneity of variances of $k$ normal populations. The exact CF of the Wilks's $\Lambda$-distribution is presented in Section~\ref{WilksCF} and the exact non-null CFs of selected related multivariate test criteria are presented in Section~\ref{NonNull}. In Section~\ref{NumInv} we introduce the Gil-Pelaez inversion and its implementation based on using the trapezoidal rule. Applicability of the numerical inversion method is illustrated in Section~\ref{Examples}, where the exact distribution is compared with some known approximations. Discussion and concluding remarks are presented in Section~\ref{Conclusions}.

\section{Characteristic function of the Bartlett's test statistic}\label{BartlettsCF}

Let $X_{l,1},\dots,X_{l,n_l}$ ($l = 1,\dots,k$) represent independent random samples from $k$ normal populations, where $X_{l,j} \sim N(\mu_l,\sigma^2_l)$ are independent normally distributed RVs with unknown means $\mu_l$ and unknown variances $\sigma^2_l$ for all $l = 1,\dots,k$ and $j = 1,\dots,n_l$. The Bartlett's $\chi^2$ test statistic (the corrected version of the log-likelihood based test statistic) and its approximate null distribution for testing homogeneity of variances of $k$ normal populations, i.e.~the hypothesis $H_0: \sigma^2_1=\cdots=\sigma^2_k$, is given by
\begin{equation}\label{eq07}
\chi^2 = \frac{\nu \log(S^2_p) - \sum_{l=1}^k \nu_l\log(S^2_l)}{1 + \frac{1}{3(k-1)}\left( \sum_{l=1}^k \frac{1}{\nu_l} - \frac{1}{\nu}\right)} \stackrel{\nu\to \infty}{\sim} \chi^2_{k-1},
\end{equation}
where $\nu_l = n_l-1$, $\nu = \sum_{l=1}^k \nu_l = N-k$ with $N = \sum_{l=1}^k n_l$, with the sample variances $S^2_l = \frac{1}{\nu_l}\sum_{j=1}^{n_l} (X_{l,j} -\bar{X}_l)^2$, for $l = 1,\dots,k$, and the pooled sample variance $S^2_p = \frac{1}{\nu}\sum_{l=1}^k \nu_l S^2_l$, where $\bar{X}_l = \sum_{j=1}^{n_l} X_{l,j}$.

The closed form expression for the exact distribution of the $\chi^2$ test statistic (\ref{eq07}) is unknown, but its distribution can be approximated by the asymptotic chi-square distribution with $k-1$ degrees of freedom, see \cite{Bartlett1937}, for more precise higher-order asymptotic approximations see \cite{Anderson2003}. However, the exact null distribution of the Bartlett's $\chi^2$ test statistic can be evaluated by numerical inversion of its CF. 

The exact distribution of the Bartlett's test statistic was studied (among others) by Glaser and Chao in \cite{Glaser1976,Glaser1976b,Chao1978}. They recognized that the null distribution of the
likelihood ratio statistic (which is functionally related to the Bartlett's $\chi^2$  test statistic) is related to the distribution of a ratio of the weighted geometric mean and the arithmetic mean of independent gamma distributed random variables. Based on that, they succeeded to derive the characteristic function of the log-likelihood ratio test statistic. However, the subsequently derived expression for PDF of the considered test statistic was expressed in a complicated and intractable form for practical purposes. In fact, they used the asymptotic expansion of the derived cumulant generating function, in order to express the probability density function of the log-likelihood ratio test statistic as an infinite linear combination of chi-square densities (with the coefficients depending on the parameters and on the complicated double sums of Bernoulli polynomials).

Here we briefly recall the basic steps of deriving the exact CF of the Bartlett's $\chi^2$ test statistic (\ref{eq07}). Let $R_w$ be a ratio of the weighted geometric mean and the arithmetic mean,
\begin{equation}\label{eq08}
R_w = \frac{G_w}{A} = \frac{\prod_{l=1}^k X^{w_l}_l}{\frac{1}{k} \sum_{l=1}^k X_l},
\end{equation}
where $w_l$ are the weights, such that $\sum_{l=1}^k w_l =1$, and $X_l \sim \mathop{\mathrm{Gamma}}(\alpha_l,\beta)$ are independent gamma distributed RVs with the shape parameters $\alpha_l$ for for $l = 1,\dots,k$ and common scale (resp.~rate) parameter $\beta$. Note that the ratio $R_w$ and the arithmetic mean $A$ are mutually independent random variables and, moreover, $R_w$  is scale invariant, i.e.~the distribution does not depend on the scale parameter $\beta$. Hence, the exact $r$th moment of $R_w$ is given by
\begin{equation}\label{eq09}
E\left(R_w^r\right) = E\left[\left( \frac{\prod_{l=1}^k X^{w_l}_l}{\frac{1}{k} \sum_{l=1}^k X_l}\right)^r\right] = \frac{\prod_{l=1}^k E\left(X^{rw_l}_l\right)}{E\left[\left( \frac{1}{k} \sum_{l=1}^k X_l\right)^r\right]} =  \frac{\prod_{l=1}^k E\left(X^{rw_l}_l\right)}{\left( \frac{1}{k}\right)^r E\left(X^r\right)},
\end{equation}
with $X = \sum_{l=1}^k X_l \sim  \mathop{\mathrm{Gamma}}(\sum_{l=1}^k \alpha_l,\beta)$. Further, by using our knowledge about the $r$th moment of the gamma distribution, i.e.
\begin{equation}\label{eq10}
E\left(Y^r\right) = \frac{\beta^r \Gamma(\alpha +r)}{\Gamma(\alpha)},
\end{equation}
for $Y \sim \mathop{\mathrm{Gamma}}(\alpha,\beta)$, we directly get the expression for the $r$th moment of $R_w$, 
\begin{equation}\label{eq11}
E\left(R_w^r\right) = k^r \frac{\Gamma\left(\sum_{l=1}^k \alpha_l \right)}{\Gamma\left(\sum_{l=1}^k \alpha_l +r\right)} \prod_{l=1}^k  \frac{\Gamma\left(\alpha_l + r w_l \right)}{\Gamma\left(\alpha_l\right)}.
\end{equation}
In general, for any log-transformed non-negative RV, say $Y = \log(X)$, its CF can be derived from the formal expression of the $r$th moment of $X$ (if it exists and is well defined also for purely imaginary \emph{order}, say $r = \mathrm{i}t$) by substituting the order $r$ with the complex variable $\mathrm{i} t$. In particular, 
\begin{equation}\label{eq12}
E\left(X^r\right) = E\left(e^{r\log(X)}\right) \Rightarrow E\left(e^{\mathrm{i} t\log(X)}\right) = \mathop{\mathrm{cf}}\nolimits_{\log(X)}(t) = \mathop{\mathrm{cf}}\nolimits_{Y}(t).
\end{equation}

By using (\ref{eq11}) and (\ref{eq12}) we get the expression for the characteristic function of $W = \log\left(R_w\right)$. In particular,
\begin{equation}\label{eq13}
\mathop{\mathrm{cf}}\nolimits_{W}(t)  =  k^{\mathrm{i}t} \frac{\Gamma\left(\sum_{l=1}^k \alpha_l \right)}{\Gamma\left(\sum_{l=1}^k \alpha_l +\mathrm{i}t\right)} \prod_{l=1}^k  \frac{\Gamma\left(\alpha_l +  \mathrm{i}w_l t\right)}{\Gamma\left(\alpha_l\right)}.
\end{equation}

Now, let $L$ denote the likelihood ratio based test statistic for testing homogeneity of variances in $k$ normal populations with unequal sample sizes, which is
the ratio of the weighted geometric mean and the weighted arithmetic mean of the sample variances $S^2_l$, 
\begin{equation}\label{eq14}
L = \frac{\prod_{l=1}^k (S_l^2)^\frac{\nu_l}{\nu}}{\sum_{l=1}^k  \frac{\nu_l}{\nu} S_l^2} = \frac{\frac{1}{k}}{\prod_{l=1}^k w_l^{w_l}} \frac{\prod_{l=1}^k X_l^{w_l}}{\frac{1}{k}\sum_{l=1}^k X_l} = \frac{1}{c_w} R_w ,
\end{equation}
where $c_w =k\prod_{l=1}^k w_l^{w_l}$ and, under null-hypothesis $H_0$, $R_w$ is defined as in (\ref{eq08}) with $w_l = \frac{\nu_l}{\nu}$ and $X_l =  \nu_l S_l^2 \stackrel{H_0}{\sim}  \sigma^2 \chi^2_{\nu_l} \equiv \mathop{\mathrm{Gamma}}\left(\frac{\nu_l}{2},\frac{1}{2\sigma^2}\right)$ for $l = 1,\dots,k$. 
Obviously, the Bartlett's $\chi^2$ statistic (\ref{eq07}) is related to the likelihood ratio based statistic $L$ given in (\ref{eq14}),
\begin{equation}\label{eq15}
\chi^2 = -\frac{\nu}{b} \log(L) = \frac{\nu\log(c_w)}{b}  - \frac{\nu}{b}\log(R_w) = \frac{c}{b}  - \frac{\nu}{b}W,
\end{equation}
where $b = 1 + \frac{1}{3(k-1)}\left( \sum_{l=1}^k \frac{1}{\nu_l} - \frac{1}{\nu}\right)$ is the Bartlett's correction factor and $c = \nu\log(c_w) = \nu\log(\frac{k}{\nu}) + \sum_{l=1}^k \nu_l\log(\nu_l)$.

Finally, by using the characteristic function of $W$ derived in (\ref{eq13}), with the parameters $\alpha_l = \frac{\nu_l}{2}$ and $w_l = \frac{\nu_l}{\nu}$ for $l=1,\dots,k$, we get the exact characteristic function of the Bartlett's $\chi^2$ statistic (\ref{eq07}),
\begin{equation}\label{eq16}
\mathop{\mathrm{cf}}\nolimits_{\chi^2}(t) = e^{\mathrm{i} \frac{c}{b} t} k^{-\mathrm{i} \frac{\nu}{b} t} \frac{\Gamma\left(\frac{\nu}{2}\right)}{\Gamma\left(\frac{\nu}{2} - \mathrm{i} \frac{\nu}{b} t\right)} \prod_{l=1}^k \frac{\Gamma\left(\frac{\nu_l}{2} - \mathrm{i} \frac{\nu_l}{b} t\right)}{\Gamma\left(\frac{\nu_l}{2}\right)}.
\end{equation}

Under the alternative hypothesis $H_A$, i.e.~when $\sigma^2_i \neq \sigma^2_j$ for some $i\neq j$, the distribution of $R_w$ specified in (\ref{eq14}) depends on 
$X_l =  \nu_l S_l^2 \stackrel{H_A}{\sim}  \sigma^2_l \chi^2_{\nu_l} \equiv \mathop{\mathrm{Gamma}}\left(\frac{\nu_l}{2},\frac{1}{2\sigma^2_l}\right)$ for $l = 1,\dots,k$ (i.e.~the independent gamma distributions of $X_l$ have different shape parameters as well as different scale parameters). The exact $r$th moment of $R_w$ and the associated non-null distribution characteristic function of $\chi^2$ test statistic is more complicated to derive, and hence, it is not presented here. However, the exact non-null distribution moments of the related likelihood ratio test statistic for testing sphericity of the multivariate distribution have been derived by Khatri and Srivastava in \cite{Khatri1971}. For more details on the non-null characteristic functions and distributions of selected multivariate test criteria see Section~\ref{NonNull}.

\section{Characteritic function of the Wilks's test statistic}\label{WilksCF}

The Wilks's test statistic is frequently used in multivariate hypothesis testing, especially with regard to different likelihood-ratio tests and multivariate analysis of variance (MANOVA). Let ${E} \sim W_{p}({n},\Sigma)$ and ${H} \sim W_{p}({q},\Sigma)$ are independent ${p}$-dimensional Wishart matrices representing the residual errors and the hypothesis model sums of squares and products matrices, with the respective degrees of freedom ${n}$ and ${q}$, such that ${n}\geq {p}$, and a common covariance matrix $\Sigma$. The Wilks's $\Lambda$ statistic and its exact null distribution is given by
\begin{equation}\label{eq17}
\Lambda = \frac{|{E}|}{|{E}+{H}|} \sim \prod_{j=1}^p B_j \equiv \Lambda(p,n,q), 
\end{equation}
where $B_j\sim \mathop{\mathrm{Beta}}\left(\frac{{n}-j+1}{2},\frac{{q}}{2}\right)$ are independent RVs with beta distributions, with specific (different) parameters for $j=1,\dots,{p}$, and by $\Lambda(p,n,q)$ we denote the Wilks's Lambda distribution with the parameters $p$, $n$, and $q$, see e.g.~\cite{Anderson2003}.

The exact Wilks's $\Lambda$ distribution with the parameters $p$ (number of variates), ${n}$ (error degrees of freedom), and ${q}$ (hypothesis degrees of freedom), have been broadly studied in statistical literature for specific parameters $p$, $q$ as well as for quite general situation with arbitrary parameters, see e.g.~\cite{Wald1941,Schatzoff1966,Pillai1969,Mathai1971,Davis1979}. In particular, Wald and Brookner in \cite{Wald1941} gave an expansion of the exact CDF of $\log (\Lambda)$ from its CF by using the method of residues, expressed in general as an infinite series expansion applicable for any grouping, Schatzoff in \cite{Schatzoff1966} considered the representation of $-\log (\Lambda)$ as a sum of independently distributed beta random variables and derived the expressions for its distribution (PDF/CDF) by taking successive convolutions. He showed how to compute numerical values of the coefficients in the derived expressions (for both the density and distribution functions) by recursive computational techniques. 
Pillai and Gupta in \cite{Pillai1969} derived explicit expressions for $p = 3,\dots,6$. Mathai and Rathie in \cite{Mathai1971} derived the exact distribution by using the technique based on the inverse Mellin transform. However, computational problems may arise in tabulating the distributions when the latter are obtained as infinite series. One possibility for overcoming the problem of a slowly convergent series is to expand the series at intermediate points, and so to approach the required percentage points by a process of 'analytic continuation'. If the distribution can be shown to satisfy a differential equation, the latter may provide a convenient tool for this process, as was suggested by Davis in \cite{Davis1979}.

However, the derived closed form expressions of the distribution functions are typically too complicated for practical purposes, especially for higher values of the parameters $p$ and $q$, and thus frequently approximated by using the well-known asymptotic approximation, $-{n}\log(\Lambda) \stackrel{{n}\to \infty}{\sim} \chi^2_{{p} {q}}$, and/or its improved (corrected) versions, see \cite{Bartlett1938} and \cite{Rao1948}, 
\begin{equation}\label{eq18}
-n\left(1 - \frac{p-q+1}{2n}\right)\log(\Lambda) \stackrel{{n}\to \infty}{\sim} \chi^2_{{p} {q}},
\end{equation}
where $\chi^2_{{p} {q}}$ represents the chi-square distribution with ${p} {q}$ degrees of freedom, or by using other known approximations, see e.g.~\cite{Fujikoshi2006,Ulyanov2006,Grilo2010,Grilo2012,Coelho2017}. 

In any case, the exact distribution of the log-transformed statistic $\lambda = -\log(\Lambda)$ can be evaluated by numerical inversion of its CF. In particular, 
\begin{eqnarray}\label{eq19}
\mathop{\mathrm{cf}}\nolimits_{\lambda}(t) = \mathop{\mathrm{cf}}\nolimits_{\log(\Lambda)}(-t) = \prod_{j=1}^{p} \mathop{\mathrm{cf}}\nolimits_{\log(B_j)}(-t) 
= \prod_{j=1}^{p}  \frac{\Gamma\left(\frac{{n}-j+1}{2} - \mathrm{i} t \right)}{\Gamma\left(\frac{{n}-j+1}{2} \right)} \frac{\Gamma\left(\frac{{n}+{q}-j+1}{2} \right)}{\Gamma\left(\frac{{n}+{q}-j+1}{2} - \mathrm{i} t \right)},
\end{eqnarray}
where $\mathop{\mathrm{cf}}\nolimits_{\log(B_j)}(t)$ denotes the CF of the log-transformed random variable $Y_j = \log(B_j)$ for $j = 1,\dots,{p}$. 

The characteristic function (\ref{eq19}) was derived by using (\ref{eq12}) and the knowledge about the $r$th moment of the beta distribution, i.e. 
\begin{eqnarray}\label{eq20}
E\left(B^r\right) = \prod_{j=0}^{r-1} \frac{\alpha+j}{\alpha+\beta+j} = \frac{\Gamma(\alpha+r)}{\Gamma(\alpha)}\frac{\Gamma(\alpha+\beta)}{\Gamma(\alpha+\beta+r)},
\end{eqnarray}
where $B\sim \mathop{\mathrm{Beta}}\left(\alpha,\beta\right)$. 

One possible application of the test statistic (\ref{eq17}) is related to the well-known likelihood ratio test (LRT) for the equality of $p$-dimensional mean vectors $\mu_l$ of $q$ normal distributions, $N_p(\mu_l,\Sigma)$ for $l=1,\dots,q$, when the common covariance matrix $\Sigma$ is assumed to be just positive-definite, but otherwise unstructured, see \cite{Anderson2003}. The null hypothesis $H_0: \mu_1 =\cdots=\mu_q$ can be tested based on $q$ independent random samples $X_{l,1},\dots,X_{l,n_l}$ with $X_{l,j}\sim N_p(\mu_l,\Sigma)$ ($l = 1,\dots,q$, $j = 1,\dots,n_l$),  by the LRT test statistic
\begin{equation}\label{eq21}
\Lambda = \frac{|E|}{|E+H|} \stackrel{H_0}{\sim} \Lambda(p,n-q,q-1),
\end{equation}
where $n=\sum_{l=1}^q n_l$, $E = \sum_{l=1}^q\sum_{j=1}^{n_l} (X_{l,j} - \bar{X}_l)(X_{l,j} - \bar{X}_l)' \sim W_{p}({n}-{q},\Sigma)$ and $H = \sum_{l=1}^q (\bar{X}_l - \bar{X})(\bar{X}_l - \bar{X})'\sim W_{p}({q}-1,\Sigma)$, with $\bar{X}_l = \frac{1}{n_l}\sum_{j=1}^{n_l} X_{l,j}$ and $\bar{X} = \frac{1}{q}\sum_{l=1}^{q} \bar{X}_l$.

In \cite{Coelho2017}, Coelho suggested similar test for cases where the common covariance matrix $\Sigma$ is restricted by some given structure, in particular the compound symmetry structure, i.e. $\Sigma = \Sigma_{CS} = (a-b) I_p + b J_p$ for some (unknown) parameters $a>0$ and $b$ such that $-\frac{a}{p-1}<b<a$. 

In such situation the null hypothesis is $H_0: \mu_1 =\cdots=\mu_q$ with assuming $\Sigma_1 = \cdots = \Sigma_q = \Sigma_{CS}$. Now, we get $E \sim W_{p}({n}-{q},\Sigma_{CS})$ and $H\sim W_{p}({q}-1,\Sigma_{CS})$. Based on that, the suggested LRT statistic which has similar structure as in (\ref{eq17}) resp.~(\ref{eq21}) and its null distribution is given by
\begin{equation}\label{eq22}
\Lambda_{CS} = \frac{a^{**}_{11} (a^{**})^{p-1}}{c^{**}_{11}(c^{**})^{p-1}}  \stackrel{H_0}{\sim} B_1 (B_2)^{{p}-1},
\end{equation}
where $a^{**} = \frac{1}{p-1}\sum_{i=2}^p a^{**}_{ii}$, $c^{**} = \frac{1}{p-1}\sum_{i=2}^p c^{**}_{ii}$, with $a^{**}_{ii}$ and $c^{**}_{ii}$ denoting the diagonal elements of the matrices $A^{**} = UEU'$  and $C^{**} = U(E+H)U'$, where $E$ and $H$ are defined as before and $U$ denotes the $(p\times p)$-dimensional Helmert matrix, and finally, $B_1\sim \mathop{\mathrm{Beta}}\left(\frac{{n}-{q}}{2},\frac{{q}-1}{2}\right)$ and $B_2 \sim  \mathop{\mathrm{Beta}}\left(\frac{({n}-{q})({p}-1)}{2},\frac{({q}-1)({p}-1)}{2}\right)$ denote two independent beta distributed random variables. 

Hence, by using (\ref{eq12}), (\ref{eq20}) and (\ref{eq22}), we get the characteristic function of the log-transformed test statistic $\lambda_{CS} = -\log(\Lambda_{CS})$,
\begin{eqnarray}\label{eq23}
\mathop{\mathrm{cf}}\nolimits_{\lambda_{CS}}(t) &=& \mathop{\mathrm{cf}}\nolimits_{\log(B_1)}(-t) \mathop{\mathrm{cf}}\nolimits_{\log(B_2)}(-(p-1)t)\cr
&=& \frac{\Gamma\left(\frac{{n}-{1}}{2}\right) \Gamma\left(\frac{{n}-{q}}{2} - \mathrm{i} t \right)}{\Gamma\left(\frac{{n}-{q}}{2}\right) \Gamma\left(\frac{{n}-{1}}{2} - \mathrm{i} t \right)}
 \frac{\Gamma\left(\frac{({n}-{1})({p}-1)}{2}\right) \Gamma\left(\frac{({n}-{q})({p}-1)}{2} - \mathrm{i}({p}-1) t \right)}{\Gamma\left(\frac{({n}-{q})({p}-1)}{2}\right) \Gamma\left(\frac{({n}-{1})({p}-1)}{2} - \mathrm{i}({p}-1) t \right)}.
\end{eqnarray}
The exact distribution of the log-transformed statistic $\lambda_{CS} = -\log(\Lambda_{CS})$ can be evaluated by numerical inversion of its CF. For more details on derivation of the test statistic and its characteristic function and alternative methods for evaluating its distribution see \cite{Coelho2017}.

\section{Characteristic functions of the non-null distributions}\label{NonNull}

In general, the exact non-null distributions of the multivariate test criteria are unknown or difficult to derive. As noted in \cite{Mathai1973}, a breakthrough in this field was possible to achieve by using special functions with matrix arguments, in particular, by using the hypergeometric functions with matrix argument or the generalized functions, such as the Meijer's $G$-function or the Fox's $H$-function, for more details see \cite{Mathai2008,Mathai2009,Muirhead2009}. In particular, the generalized hypergeometric function with matrix argument is defined by
\begin{eqnarray}\label{eq24}
_pF_q\left(a_1,\dots,a_p;b_1,\dots,b_q \,|\, X \right) = \sum_{k=1}^\infty\sum_\kappa \frac{(a_1)_\kappa \cdots (a_p)_\kappa}{k!(b_1)_\kappa \cdots (b_q)_\kappa}C_\kappa(X)
\end{eqnarray}
where $p\geq 0$ and $q \geq 0$ are integers, and $X$ is $n\times n$ symmetric matrix with eigenvalues $x_1, x_2,\dots, x_n$, $\kappa = (\kappa_1, \kappa_2,\dots)$ is a partition of $k$, 
$(a)_\kappa$ and $(b)_\kappa$ represent the generalized Pochhammer symbols, and $C_\kappa(X)$ is the Jack function --- a symmetric, homogeneous polynomial of degree $|\kappa|$ in the eigenvalues $x_1, x_2, \dots, x_n$ of $X$. For more details and strategies for efficient computation of the generalized hypergeometric function see \cite{Koev2006,KoevMHF2008}. Buttler and Wood in \cite{Butler2002,Butler2005} suggested efficient Laplace approximations for two functions of matrix argument: the Type I confluent hypergeometric function, $_1F_1\left(a;b \,|\, X\right)$, and the Gauss hypergeometric function, $_2F_1\left(a,b; c \,|\, X\right)$.

In special cases it is possible to evaluate the moments of the statistics under consideration. For example, see \cite{Mathai1973}, the $r$th moment of Wilks's generalized variance $|{S}|$, where ${S}$ is a non-central Wishart distribution with $n$ degrees of freedom and the parameters $\Sigma$ (covariance matrix) and $\Omega$ (non-centrality matrix), ${S}\sim W_{p}({n},\Sigma,\Omega)$, is 
\begin{eqnarray}\label{eq25}
E\left(|S|^r\right) = \frac{\Gamma_p(\frac{n}{2}+r)}{\Gamma_p(\frac{n}{2})} |2\Sigma|^r \exp\left(-\mathop{\mathrm{trace}}(\Omega)\right) \,_1F_1\left(\frac{n}{2}+r;\frac{n}{2} \,|\, \Omega \right),
\end{eqnarray}
where $\Gamma_p(a)$ denotes the multivariate gamma function, 
\begin{eqnarray}\label{eq26}
\Gamma_p(a) = \pi^{\frac{p(p-1)}{2}} \prod_{j=1}^p\Gamma\left(a - \frac{j-1}{2}\right).
\end{eqnarray}
By using (\ref{eq12}), the characteristic function of $W = -\log(|S|)$ is
\begin{eqnarray}\label{eq27}
\mathop{\mathrm{cf}}\nolimits_{W}(t) = \frac{\Gamma_p(\frac{n}{2}-\mathrm{i}t)}{\Gamma_p(\frac{n}{2})} |2\Sigma|^{-\mathrm{i}t} \exp\left(-\mathop{\mathrm{trace}}(\Omega)\right) \,_1F_1\left(\frac{n}{2}-\mathrm{i}t;\frac{n}{2} \,|\, \Omega \right),
\end{eqnarray}
and hence, the required PDF/CDF/QF can be computed straightforwardly through numerical inversion of the CF (\ref{eq27}).

In the normal theory of testing hypotheses on regression coefficients the Wilks's $\Lambda$ test criterion is specified in (\ref{eq17}) with the $(p\times p)$-matrices $H$ and $E$. In general, $H$ has a non-central Wishart distribution with $q$ degrees of freedom, the covariance matrix $\Sigma$, and the matrix of non-centrality parameters $\Omega = \frac{1}{2}MM'\Sigma^{-1}$, where $M = E(X)$ is the true expectation of $X$ (if the null hypothesis is not true), where $X$ is such that $H = XX'$, i.e.~$H\sim W_p(q,\Sigma,\Omega)$. The matrix $E$ has a central Wishart distribution with $n$ degrees of freedom and a common covariance matrix $\Sigma$, $E\sim W_p(n,\Sigma)$.

Then, the non-null characteristic function of $\lambda = -\log(\Lambda)$, derived from the $r$th non-null moment of $\Lambda$, as specified in \cite{Constantine1963}, is
\begin{eqnarray}\label{eq28}
\mathop{\mathrm{cf}}\nolimits_{\lambda}(t) = \frac{\Gamma_p\left(\frac{{n}}{2} - \mathrm{i} t \right)}{\Gamma_p\left(\frac{{n}}{2} \right)} \frac{\Gamma_p\left(\frac{{n}+{q}}{2} \right)}{\Gamma_p\left(\frac{{n}+{q}}{2} - \mathrm{i} t \right)} \,_1F_1\left(-\mathrm{i}t;\frac{{n}+{q}}{2} - \mathrm{i} t \,|\, -\Omega \right).
\end{eqnarray}
Note that under the null hypothesis, i.e.~if $\Omega = 0$, the characteristic function (\ref{eq28}) coincides with (\ref{eq19}). For more details see also \cite{Muirhead2009,Butler2005}.

Similarly the test criterion for testing equality of covariances of two $p$-dimensional multivariate normal populations of size $N_1$ and $N_2$, based on the test statistic
\begin{eqnarray}\label{eq29}
\Lambda_2 = \frac{|A_1|^{\frac{n_1}{n}} |A_2|^{\frac{n_2}{n}}}{|A_1+A_2|},
\end{eqnarray}
where $n_1=N_1-1$, $n_2=N_2-1$, $n = n_1+n_1$, $A_1 = \sum_{i=1}^{N_1} (X_i-\bar{X})(X_i-\bar{X})'$ and $A_2 = \sum_{i=1}^{N_2} (Y_i-\bar{Y})(Y_i-\bar{Y})'$. 
The non-central distribution of $\Lambda_2$ under $H_A: \Sigma_1\neq\Sigma_2$ is determined by the parameters $p$, $n_1$, $n_2$ and the eigenvalues $\delta_1,\dots, \delta_p$ of the matrix $\Delta = \Sigma_1\Sigma_2^{-1}$. In particular, the non-null characteristic function of $\lambda_2 = -\log(\Lambda_2)$ derived from the $r$th non-null moment of $\Lambda_2$, as specified in \cite{Constantine1963}, is
\begin{eqnarray}\label{eq30}
\mathop{\mathrm{cf}}\nolimits_{\lambda_2}(t) 
= \frac{\Gamma_p\left(\frac{{n}}{2}\right)}{\Gamma_p\left(\frac{{n}}{2}(1-\frac{2\mathrm{i} t}{n})\right)}
\frac{\Gamma_p\left(\frac{{n_1}}{2}(1-\frac{2\mathrm{i} t}{n})\right)}{\Gamma_p\left(\frac{{n_1}}{2} \right)}
\frac{\Gamma_p\left(\frac{{n_2}}{2}(1-\frac{2\mathrm{i} t}{n})\right)}{\Gamma_p\left(\frac{{n_2}}{2} \right)}
 |\Delta|^{-\frac{n_1\mathrm{i}t}{n}}  \,_2F_1\left(-\mathrm{i}t, \frac{{n_1}}{2}\left(1-\frac{2\mathrm{i} t}{n}\right);\frac{{n}}{2}\left(1-\frac{2\mathrm{i} t}{n}\right)\,|\, I_p -\Delta \right).
\end{eqnarray}

The required PDF/CDF/QF of the  non-null distributions can be computed by numerical inversion of their CFs, (\ref{eq27}) (\ref{eq28}) and (\ref{eq30}), by using algorithms for computing the generalized hypergeometric functions with matrix argument, e.g., as suggested in \cite{KoevMHF2008}, or by using suitable approximations, see e.g.~\cite{Butler2002}. In fact, evaluation of the generalized hypergeometric functions with matrix argument is still a big challenge and numerical precision and efficiency of the computation strongly depends on the quality of the available algorithms. 

In general, once the non-null distribution moments of the considered multivariate test statistic are available the characteristic function of the log-transformed statistic can be derived and the numerical inversion of the CF can be applied to evaluate the exact PDF/CDF and the quantiles. The non-null moments of the multivariate test criteria have been broadly discussed in statistical literature, for more particular cases see e.g.~\cite{Anderson2003,Khatri1971,Mathai1973,Constantine1963}. However, for many important test criteria the characteristic functions or the non-null moments are still not available or difficult to compute. These are open problems for further research. 

\section{Methods and algorithms for numerical inversion of the characteristic functions}\label{NumInv}

Let $Y$ denotes the continuous univariate RV with its PDF $\mathop{\mathrm{pdf}}\nolimits_{Y}(y)$. Recall that the CF of the distribution of $Y$, given by the Fourier transform of its PDF, is defined as
\begin{equation}\label{eq31}
\mathop{\mathrm{cf}}\nolimits_{Y}(t)   = \mathop{\cal F}\Big(\mathop{\mathrm{pdf}}\nolimits_{Y}(\cdot)\Big)(t)
= \mathop{E}\left[e^{\mathrm{i}tY}\right]  = \int_{-\infty}^\infty e^{\mathrm{i}ty} \mathop{\mathrm{pdf}}\nolimits_{Y}(y) \,dy .
\end{equation}
Conversely, the PDF of $Y$ is the inverse Fourier transform of its CF,
\begin{eqnarray}\label{eq32}
\mathop{\mathrm{pdf}}\nolimits_{Y}(y) 
= \mathop{\cal F}\nolimits^{-1}\Big(\mathop{\mathrm{cf}}\nolimits_{Y}(\cdot)\Big)(y)
= \frac{1}{2\pi}\int_{-\infty }^{\infty} e^{-\mathrm{i}ty}\mathop{\mathrm{cf}}\nolimits_{Y}(t)\,dt
= \frac{1}{\pi}\int_0^\infty \Re\left(e^{-\mathrm{i}ty}\mathop{\mathrm{cf}}\nolimits_{Y}(t) \right)\,dt,
\end{eqnarray}
where $\Re(z)$ denotes the real part of $z$. Further, CDF of $Y$ can be computed from the analytic CF via the Hilbert transform,
\begin{equation}\label{eq33}
\mathop{\mathrm{cdf}}\nolimits_{Y}(y) 
= \mathop{\cal F}\Big(\mathop{\mathbf 1}\nolimits_{(-\infty,y)}\mathop{\mathrm{pdf}}\nolimits_{Y}(\cdot)\Big)(0)
= \frac{1}{2} - \frac{\mathrm{i}}{2}\mathop{\cal H}\Big(e^{-\mathrm{i}y\cdot} \mathop{\mathrm{cf}}\nolimits_{Y}(\cdot)\Big)(0),
\end{equation}
for more details see, e.g., \cite{feng2013inverting}.

Computing the (inverse) Fourier transform numerically is a well-known problem, frequently connected with the problem of computing integrals of highly oscillatory (complex) functions. The problem was studied for a long time in general, but also with focus on specific applications, see, e.g., \cite{asheim2013complex,levin1996fast,milovanovic1998numerical,sidi1982numerical,sidi1988user,sidi2012user}. In particular, the methods suggested for inverting the characteristic function for obtaining the probability distribution function include \cite{abate1992fourier,shephard1991characteristic,waller1995obtaining,zielinski2001high,strawderman2004computing}. 

If CF is absolutely integrable over $(-\infty,\infty)$, Gil-Pelaez in \cite{GilPelaez1951} derived the inversion formula which require integration of a real-valued function, only. In particular,  
\begin{eqnarray}\label{eq34}
\mathop{\mathrm{cdf}}\nolimits_{Y}(y) 
 = \frac{1}{2}-\frac{1}{2\pi}\int_0^\infty \frac{e^{-\mathrm{i}ty} \mathop{\mathrm{cf}}\nolimits_{Y}(t) -e^{\mathrm{i}ty} \mathop{\mathrm{cf}}\nolimits_{Y}(-t)}{\mathrm{i} t} \,dt
 = \frac{1}{2}-\frac{1}{\pi}\int_0^\infty \Im\left(\frac{e^{-\mathrm{i}ty}\mathop{\mathrm{cf}}\nolimits_{Y}(t) }{t} \right)\,dt,
\end{eqnarray}
where $\Im(z)$ denotes the imaginary part of $z$. The Gil-Pelaez inversion formulae can be evaluated by using a simple trapezoidal rule:
\begin{equation}\label{eq35}
	\mathop{\mathrm{pdf}}\nolimits_{Y}(y) 
\approx	\frac{\delta_t}{\pi} \sum_{j=0}^N w_j \Re\left(e^{-\mathrm{i}t_jy}\mathop{\mathrm{cf}}\nolimits_{Y}(t_j)  \right),
\end{equation}
\begin{equation}\label{eq36}
\mathop{\mathrm{cdf}}\nolimits_{Y}(y) 
	\approx	\frac{1}{2}-\frac{\delta_t}{\pi} \sum_{j=0}^N w_j  \Im\left(\frac{e^{-\mathrm{i}t_jy}\mathop{\mathrm{cf}}\nolimits_{Y}(t_j) }{t_j} \right),
\end{equation}
where
\begin{itemize}\itemsep-1pt
  \item $N$ is sufficiently large integer,
	\item the optimum discretization step is $\delta_t = \frac{2\pi}{B-A}$, where $(A,B)$ is the domain of $Y$, 
	\item if not known explicitly or the distribution limits are infinite, $(A,B)$ can be approximated by a sufficiently large interval covering large part of the distribution domain, e.g., by using the six-sigma-rule: $(A,B) = \mathop{\mathrm{mean}}(Y)\mp 6\mathop{\mathrm{std}}(Y)$,
	\item $w_j$, $j = 0,\dots,N$, are the quadrature weights ($w_0=w_N=\frac{1}{2}$, otherwise $w_j=1$),  
	\item $t_j$, $j = 0,\dots,N$, are the equidistant nodes from $(0,T)$, where $T = N\delta_t$,
	\item The total approximation error (i.e.~the truncation error and the discretization error) can be controlled by proper selection of $(A,B)$ used for setting the step $\delta_t = \frac{2\pi}{B-A}$ and selection of sufficiently large $N$, such that the integrand in (\ref{eq31}) is sufficiently small for large arguments $t$, i.e.~$|f(t)|<\varepsilon$ for all $t> T$.	
\end{itemize}

However, numerical algorithms for computing the distribution function by numerical inversion from the characteristic function are still missing in the standard statistical packages, like e.g.~SAS, R and/or MATLAB.

In order to illustrate the suggested approach, a possible alternative is to use the characteristic functions toolbox developed by the author and available at GitHub, see \cite{CharFunTool}. \texttt{CharFunTool} is a (still growing) MATLAB repository of characteristic functions and tools for their combinations and numerical inversion. The toolbox comprises different inversion algorithms, including those based on simple trapezoidal quadrature rule for computing the integrals defined by the Gil-Pelaez formulae, and/or based on using the fast Fourier transform (FFT) algorithm for computing the Fourier transform integrals, see e.g.~\cite{Carr1999,Chourdakis2004,Huerlimann2013}, as well as the algorithm for computing  non-negative continuous distributions by using the method suggested by Bakhvalov and Vasileva in \cite{Bakhvalov1968}. The method was suggested for computing the oscillatory Fourier integrals based on approximation of the integrand function by the Fourier-Legendre series expansion, and observation that Fourier transform of the Legendre polynomials is related to the Bessel $J$ functions. For more details see also \cite{Evans1999}.

\begin{figure}[t]
\begin{lstlisting}
  % Computing the exact distribution of the Bartlett's test statistic (7)
	k            = 15;                                        % number of normal populations 
	nu_l         = [1 1 1 1 1 2 2 2 2 2 3 3 3 3 3];           % sample degrees of freedom
	nu           = sum(nu_l);                                 % total degrees of freedom 
	alpha{1}     = nu_l/2;	                                  % alpha_l parameters
	weight{1}    = alpha{1}/sum(alpha{1});                    % weights_l
	c            = nu * log(k * prod(weight{1}.^weight{1}));  % coefficient c
	b            = 1 + 1/(3*(k-1))*(sum(1./nu_l) - 1/nu);     % Bartlett's correction b
	shift        = c/b;
	coef         = -nu/b;
	
	% Characteristic function of the Bartlett's test statistic (16)
	cf_logR      = @(t) cf_LogRV_MeansRatioW(t,k,alpha,weight,coef);
	cf           = @(t) exp(1i*t*shift) .* cf_logR(t);
	
	% Evaluate the distribution function by using the algorithm cf2DistGP
	x            = linspace(0,40);
	prob         = [0.9 0.95 0.99];
	options.xMin = 0;
	result       = cf2DistGP(cf,x,prob,options);
\end{lstlisting}
\includegraphics[width = 0.33\textwidth]{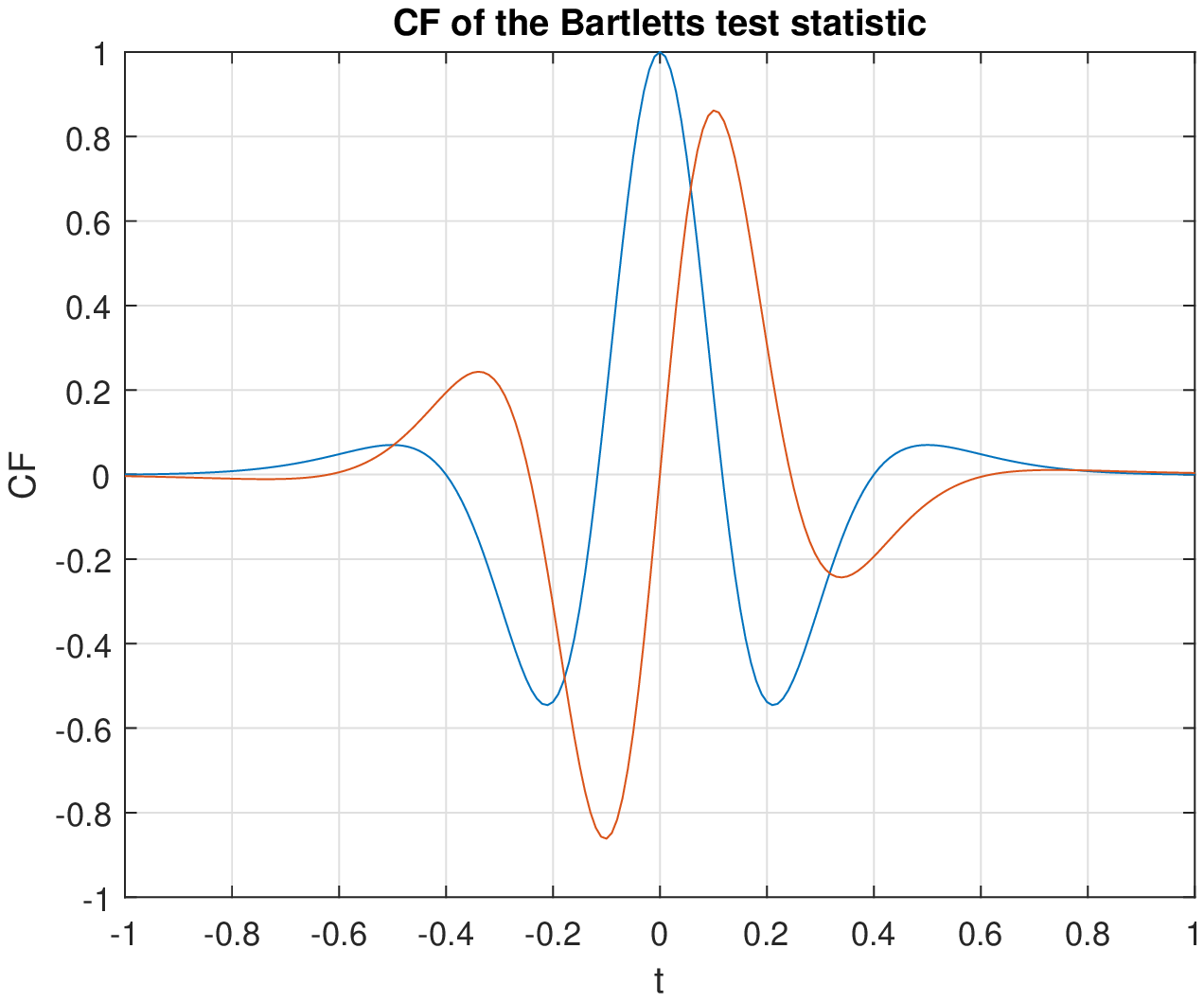}
\includegraphics[width = 0.33\textwidth]{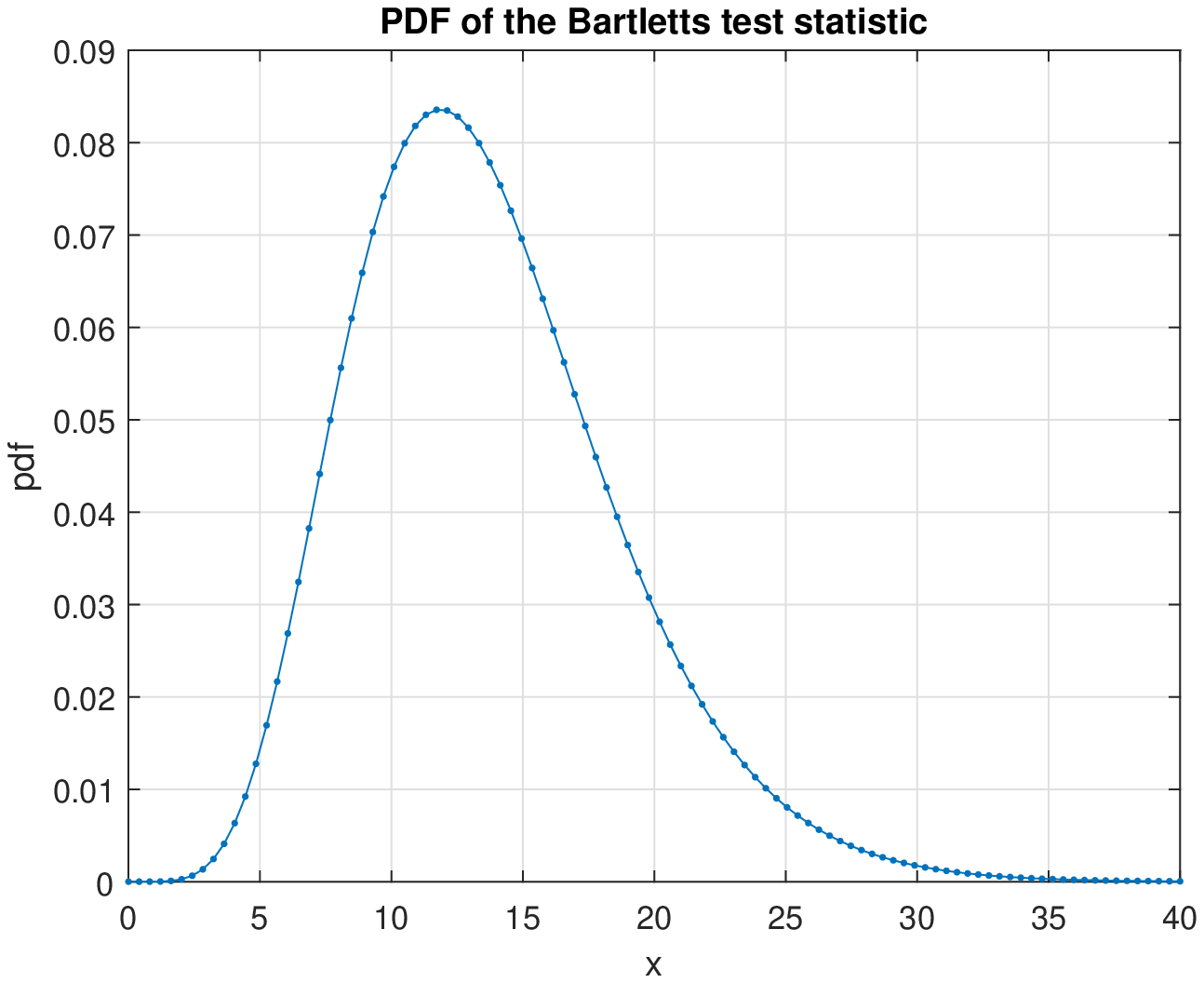}
\includegraphics[width = 0.33\textwidth]{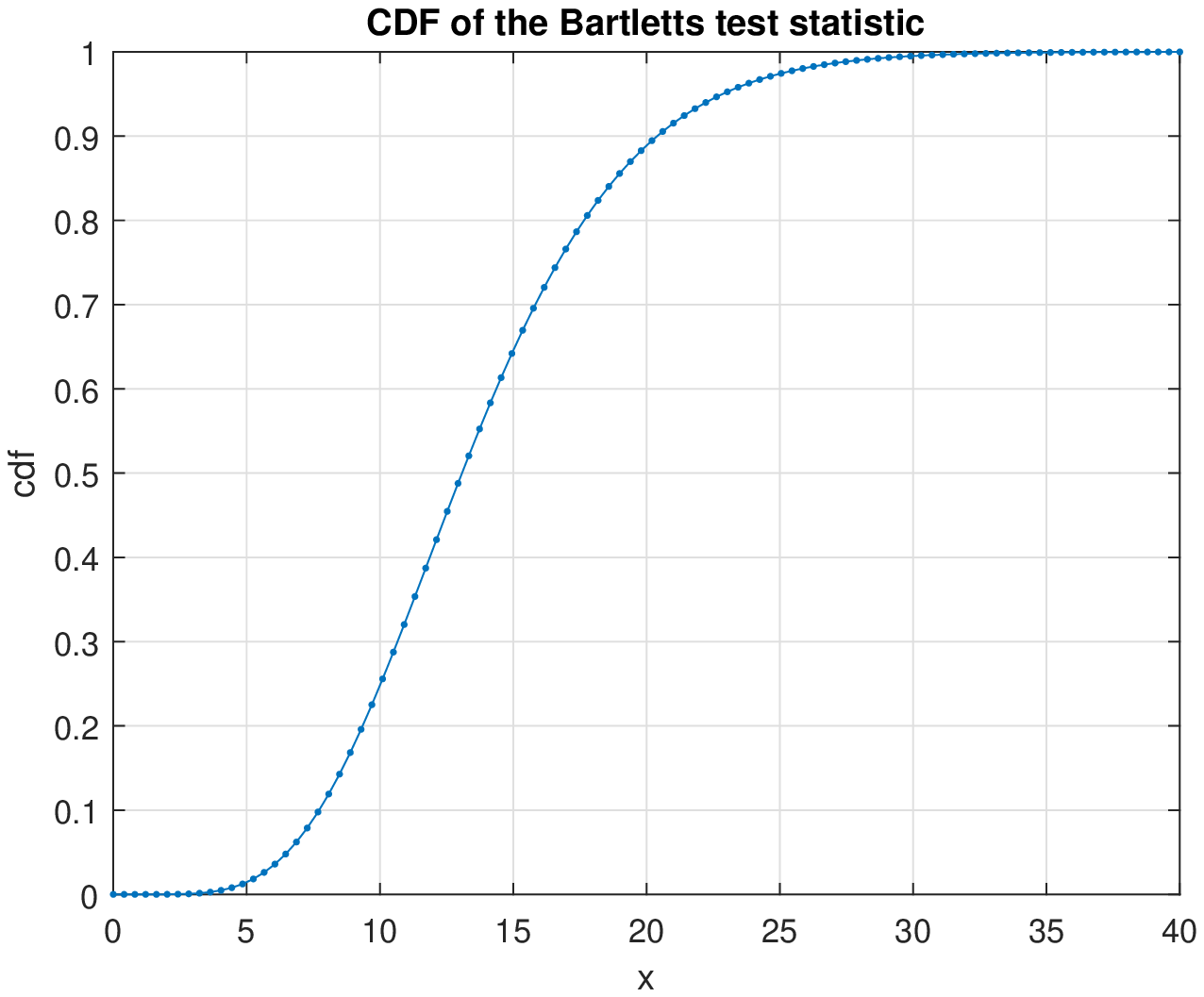}
\caption{(a) MATLAB code used to evaluate the characteristic function and the distribution functions (PDF/CDF) of the Bartlett's test statistic (\ref{eq07}). (b) Plots of the characteristic function (CF) with its real (blue) and imaginary part (red), probability density function (PDF) and the cumulative distribution function (CDF)  of the Bartlett's test statistic for testing homogeneity of $k=15$ normal populations with unequal sample sized specified by the parameters (degrees of freedom) $\nu_l$, $l=1,\dots,k$.}
\label{fig01}
\end{figure}

\section{Numerical examples}\label{Examples}

Here we illustrate the suggested approach for evaluation of the distribution (PDF/CDF) of selected test statistic by numerical inversion of their characteristic functions we present two simple examples computed by the tools and algorithms available at the MATLAB toolbox \texttt{CharFunTool}, see \cite{CharFunTool}. For more details and examples 
we recommend to check the \texttt{CharFunTool} web page and taking a look at the algorithm help files and the Examples collection.

\subsection{Computing the exact distribution of the Bartlett's test statistic}

Let us consider the exact distribution of the Bartlett's test statistic for testing homogeneity of variances of $k$ normal populations with unequal sample sizes, given by (\ref{eq07}) and specified by its characteristic function (\ref{eq16}). Let us consider the following specific parameters: 
\begin{itemize}\itemsep-1pt
\item $k=15$, number of normal populations;
\item $\nu_l \in \{1, 1, 1, 1, 1, 2, 2, 2, 2, 2, 3, 3, 3, 3, 3\}$, samples degrees of freedom $\nu_l$, $l=1,\dots,k$, with the total sum of the degrees of freedom $\nu = \sum_{l=1}^k \nu_l = 30$. 
\end{itemize}
Hence, the Bartlett's correction factor is $b = 1 + \frac{1}{3(k-1)}\left( \sum_{l=1}^k \frac{1}{\nu_l} - \frac{1}{\nu}\right) = 1.2175$ and the coefficient $c = \nu\log(\frac{k}{\nu}) + \sum_{l=1}^k \nu_l\log(\nu_l) = 2.6162$. The MATLAB code to evaluate the characteristic function and the exact distribution functions of the Bartlett's test statistic is presented in Figure~\ref{fig01}, together with the plotted graphs of the computed CF/PDF/CDF. 

For specified probabilities $0.9$, $0.95$ and $0.99$ the exact computed quantiles are $q_{0.9} =20.3969$, $q_{0.95}= 22.8508$, and $q_{0.99}=27.9221$, respectively. On the other hand, the approximate quantiles, computed from the approximate $\chi^2_{k-1}$ distribution with $k=15$, as specified in (\ref{eq07}), are $\tilde{q}_{0.9} = 21.0641$,  $\tilde{q}_{0.95} = 23.6848$, and    $\tilde{q}_{0.99} = 29.1412$, respectively.

\begin{figure}[t]
\begin{lstlisting}
    % Exact distribution of the test statistic lambda = -log(Lambda) for testing equality of p-dimensional 
    % mean vectors of q normal distributions with common unstructured covariance matrix, see (21)

    p    = 10;       % dimension of the normal populations
    q    = 7;        % number of normal populations
    n    = 30;       % total number of samples
    
    % Characteristic function of the statistic lambda = -log(Lambda) with unstructured covariance matrix, (19)
    cf_UN = @(t) cf_LogRV_WilksLambda(t,p,n-q,q-1,-1);

    % Evaluate the distribution of -log(Lambda) by using the cf2DistGP 
    x            = linspace(0,6)';
    prob         = [0.9 0.95 0.99];
    options.xMin = 0;
    result_UN    = cf2DistGP(cf_UN,x,prob,options);
		
    % Exact distribution of the test statistic lambda_CS = -log(Lambda_CS) for testing equality of  
    % p-dimensional mean vectors of q normal distributions with common covariance matrix with compound 
    % symmetry structure, see (22)
    
    % Characteristic function of the Wilks's test statistic with compound symmetry structure (23)
    cf_CS     = @(t) cf_LogRV_Beta(t,(n-q)/2,(q-1)/2,-1) .* ...
                     cf_LogRV_Beta(t,(p-1)*(n-q)/2,(p-1)*(q-1)/2,-(p-1));

    % Evaluate the distribution of -log(Lambda_CS) by using the cf2DistGP 
    result_CS = cf2DistGP(cf_CS,x,prob,options);
\end{lstlisting}
\includegraphics[width = 0.33\textwidth]{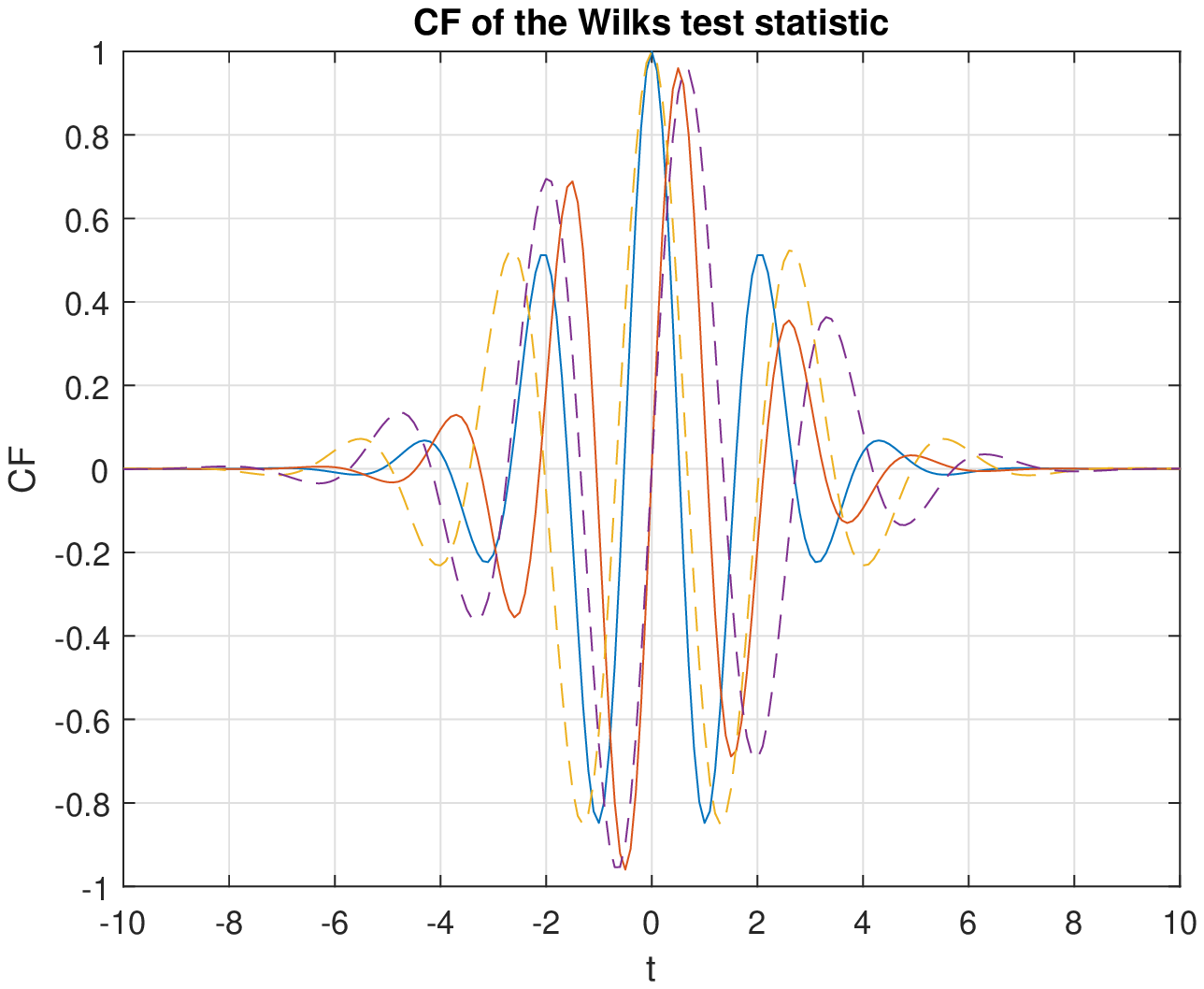}
\includegraphics[width = 0.33\textwidth]{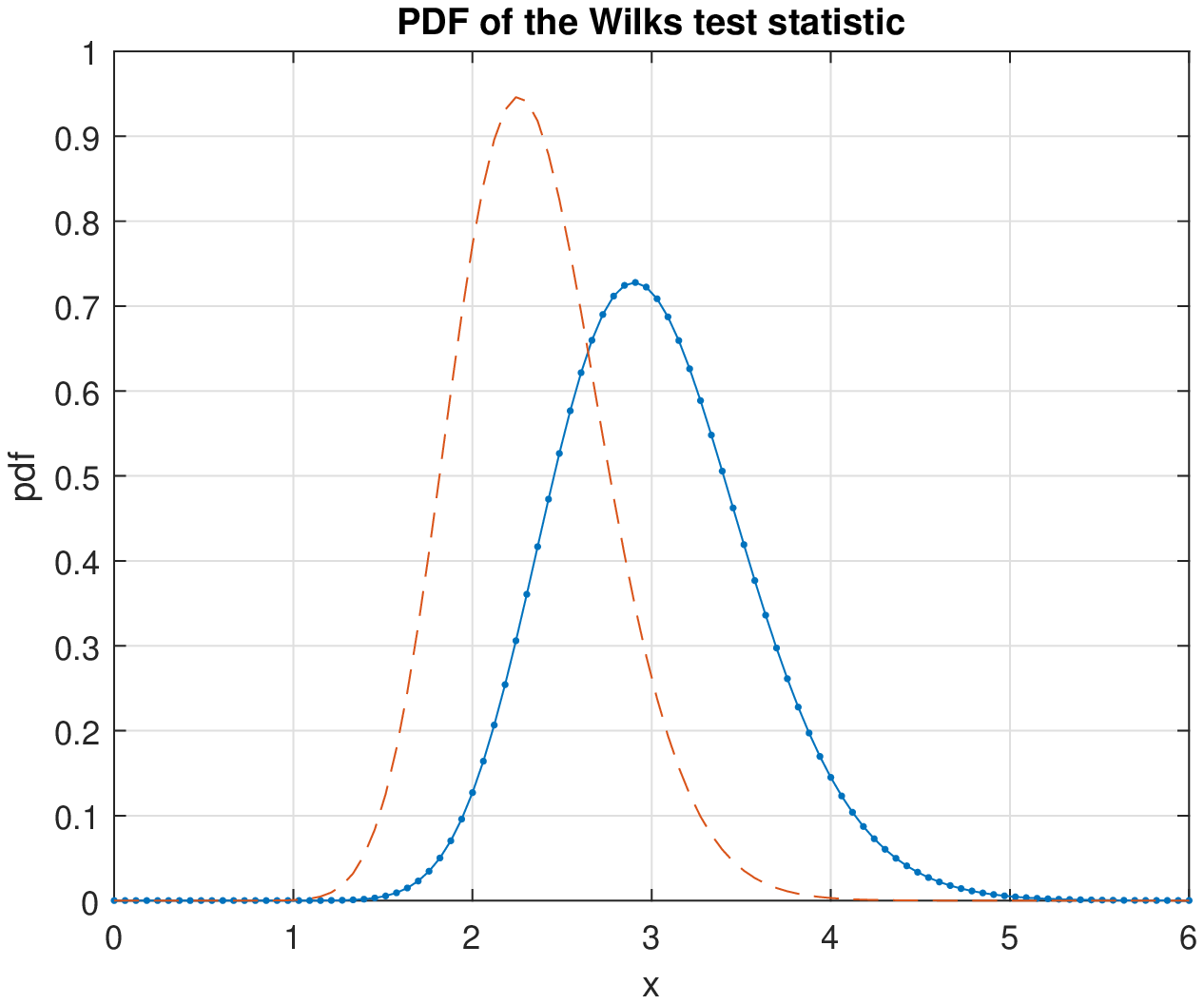}
\includegraphics[width = 0.33\textwidth]{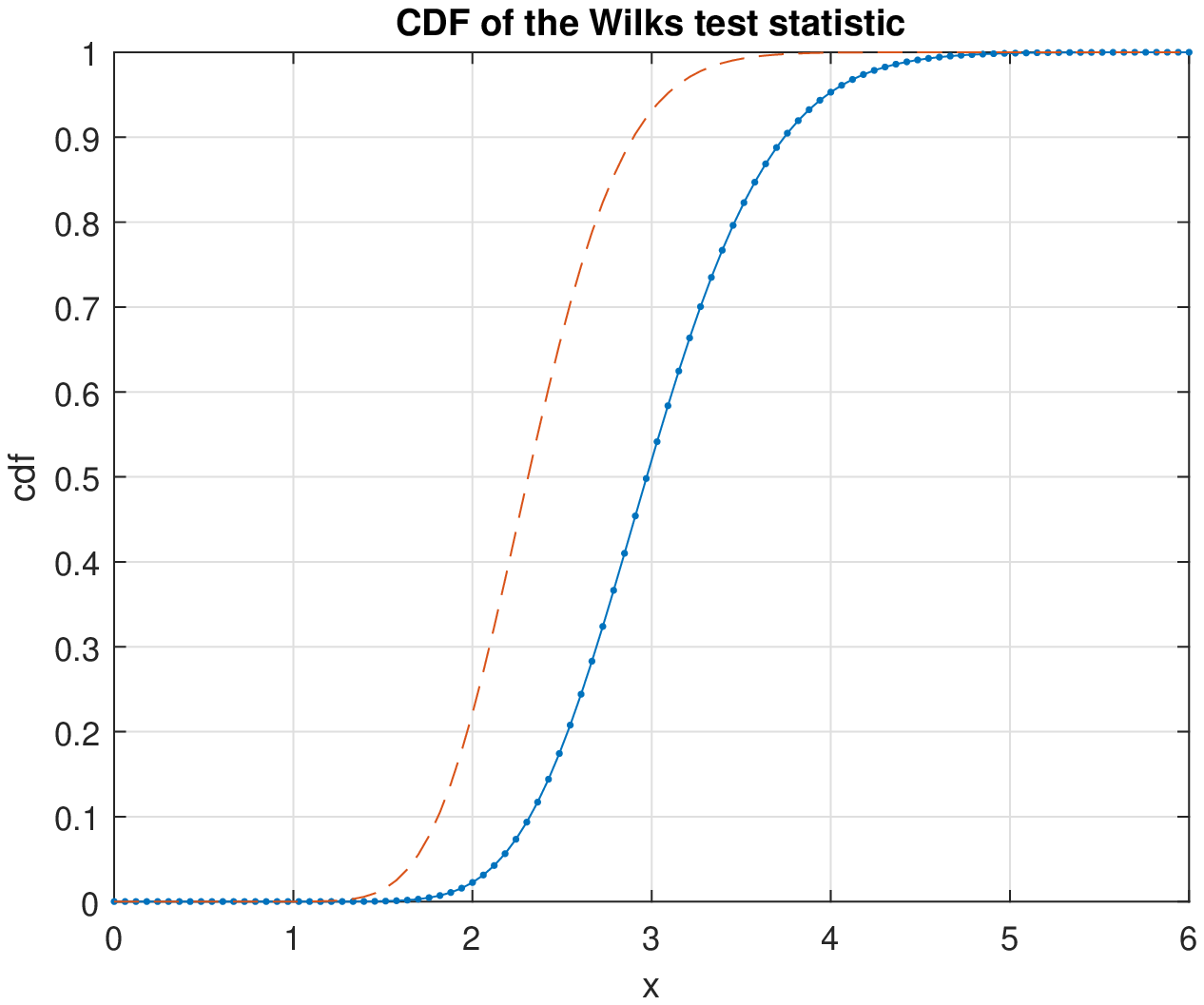}
\caption{(a) MATLAB code used to evaluate the characteristic functions and the distribution functions (PDF/CDF) of the log-transformed Wilks's test statistics, see (\ref{eq21}) when assuming the  unstructured (UN) common covariance matrix, and (\ref{eq22}) when assuming the compound symmetry (CS) of the common covariance matrix. (b) Plots of the characteristic functions (CF), probability density functions (PDF) and the cumulative distribution functions (CDF) of the log-transformed test statistics, $\lambda = -\log(\Lambda)$, with specified parameters (solid lines depict the functions for the assumed UN covariance structure, dashed lines depict the functions for the assumed CS covariance structure).}
\label{fig02}
\end{figure}

\subsection{Computing the exact distribution of the log-transformed Wilks's test statistic}
Here we consider and compare the exact distributions of the log-transformed LRT statistics, $\lambda = -\log(\Lambda)$, for testing equality of the mean vectors of $q$ normal populations, when the common covariance matrix is assumed to be unstructured (just positive-definite) and when the common covariance matrix is assumed to have given structure, in particular the compound symmetry structure.  Let us consider the following specific parameters: 
\begin{itemize}\itemsep-1pt
\item $p=10$, dimension of the normal populations;
\item $q=7$, number of normal populations;
\item $n=30$, total number of samples.
\end{itemize}
 The MATLAB code to evaluate the characteristic functions and the exact distribution functions (PDF/CDF) of the test statistics  $\lambda = -\log(\Lambda)$ under both assumed covariance structures is presented in Figure~\ref{fig02} together with the plotted graphs of the computed CF/PDF/CDF. 

Notice the apparent difference of the null distributions of the log-transformed LRT statistics for testing equality of mean vectors of the $q$ normal populations, with otherwise equal parameters, when the assumed structure of the common covariance matrix is different (unstructured vs.~compound symmetry).

\section{Conclusions}\label{Conclusions}
In general, evaluation of the exact distribution function based on numerical inversion of its characteristic function is a convenient method when the characteristic function is known apriori and is such that it can be easily numerically evaluated by the available algorithms or if the statistic under consideration is a linear combination of independent random variables with known and simple characteristic functions.

In this paper we have presented several standard test statistics used in multivariate analysis for which the exact null distribution is difficult to express analytically but its characteristic function is known and can be computed easily in standard software packages. Frequently, such distribution functions are usually approximated by using results of the asymptotic theory, or by using other available small sample approximation/correction methods, and frequently by using computer intensive simulation methods. 
Here we advocate to use the method based on numerical inversion of the characteristic functions. However, numerical algorithms for computing and combining more complicated characteristic functions and for computing the distribution function by numerical inversion from the characteristic function are still missing in the standard statistical packages, like e.g.~SAS, R and/or MATLAB. As a possible alternative and a starting point for further development here we present a MATLAB toolbox developed by the author and freely available at the GitHub, \url{https://github.com/witkovsky/CharFunTool}. Further research is necessary for deriving the characteristic functions of the non-null distributions, efficient algorithms for computing the special functions (of matrix and complex argument) required for evaluation of complicated characteristic functions, as well as development of the more advanced algorithms (efficient and precise) for numerical inversion of the characteristic functions.

\section*{Acknowledgment}
The work was supported by the Slovak Research and Development Agency, project APVV-15-0295, and by the Scientific Grant Agency VEGA of the Ministry of Education of the Slovak Republic and the Slovak Academy of Sciences, project VEGA 2/0047/15. 



\end{document}